\documentclass[sn-mathphys]{sn-jnl}

\jyear{2021}%

\theoremstyle{thmstyleone}%
\theoremstyle{thmstyletwo}%

\theoremstyle{thmstylethree}%

\usepackage{verbatim}
\begin{document}

\title[Deepsea: A Meta-ocean Prototype for Undersea Exploration ]{Deepsea: A Meta-ocean Prototype for Undersea Exploration}


\author[1]{\fnm{Jinyu} \sur{Li}}\email{lijinyu7575@163.com}

\author[2]{\fnm{Ping} \sur{Hu}}\email{pihu@cs.stonybrook.edu}

\author[3]{\fnm{Weicheng} \sur{Cui}}\email{cuiweicheng@westlake.edu.cn}
\author[1]{\fnm{Tianyi} \sur{Huang}}\email{huangtianyi@westlake.edu.cn}

\author*[1]{\fnm{Shenghui} \sur{Cheng}}\email{chengshenghui@westlake.edu.cn}

\affil*[1]{\orgdiv{Research Center for Industries of the Future and School of Engineering}, \orgname{Westlake university}, \orgaddress{\city{Hangzhou}, \country{China}}}

\affil[2]{\orgdiv{Department of Computer Science}, \orgname{Stony Brook University}, \orgaddress{\city{Stony Brook}, \country{USA}}}

\affil[3]{\orgdiv{Key Laboratory of Coastal of Environment and Resources of Zhejiang Province}, \orgname{Westlake university}, \orgaddress{\city{Hangzhou}, \country{China}}}


\abstract{
Metaverse has attracted great attention from industry and academia in recent years. 
Metaverse for the ocean (Meta-ocean) is the implementation of the Metaverse technologies in virtual emersion of the ocean which is beneficial for people yearning for the ocean.
It has demonstrated great potential for tourism and education with its strong immersion and appealing interactive user experience. However, quite limited endeavors have been spent on exploring the full possibility of Meta-ocean, especially in modeling the movements of marine creatures. 
In this paper, we first investigate the technology status of Metaverse and virtual reality (VR) and develop a prototype that builds the Meta-ocean in VR devices with strong immersive visual effects. Then, we demonstrate a method to model the undersea scene and marine creatures and propose an optimized path algorithm based on the Catmull-Rom spline to model the movements of marine life. Finally, we conduct a user study to analyze our Meta-ocean prototype. This user study illustrates that our new prototype can give us strong immersion and an appealing interactive user experience.
}

\keywords{Metaverse, Meta-ocean, virtual reality, marine reconstruction}



\maketitle




\maketitle

\section{Introduction}\label{sec1}

Metaverse is gaining increasing attention from researchers, developers, and people from various industry area.
Combining the physical reality and digital virtuality~\cite{18,Cheng2023}, metaverse brings infinite possibility and extensibility to scientific research, entertainment, education, etc~\cite{encyclopedia2010031}.
The concept of the Metaverse was first proposed in a piece of scientific fiction named Snow Crash in 1992~\cite{1}. As a computer-generated universe, metaverse has been defined through vastly diversified concepts~\cite{All}, such as lifelogging~\cite{2}, collective space in virtuality~\cite{3}, embodied Internet~\cite{4}, and Omniverse~\cite{6}.

Metaverse is a synthesis of a variety of techniques, including augmented reality (AR), mixed reality (MR), virtual reality (VR), high-speed networks, and edge computing~\cite{All}, among which VR techology~\cite{1960} has become a trendy technology in recent years\cite{cheng2022roadmap,huang2023roadmap}. 
VR technology that simulates a virtual environment immersing users to have the feeling of "being there"~\cite{Wohlgenannt2020VirtualR}, and the feeling of immersion must be supported by the proper software and hardware. 
With the advances in computer hardware and computing capability, the devices involved in VR environments have become versatile, ranging from single head-mounted displays to several complementary devices such as the EyePhone (a head-mounted display), the AudioSphere (a sound system), and the DataGlove and DataSuit (for measuring movements)~\cite{799723,Raitt1991TheEL,Wohlgenannt2020VirtualR}.


The early VR technology was mainly used for military training, flight simulation, medicine~\cite{MUHANNA2015344}, and so on by professional departments due to its high-cost hardware limitation. 
Thanks to hardware improvement, commercial VR devices such as Oculus Rift and HTC VIVE, had triggered the diffusion of VR technology into private households~\cite{Wohlgenannt2020VirtualR}.
The applications of VR technology have gradually expanded into various fields, such as the retail industry, energy industry, education industry, and tourism industry.
For example, in the retail industry, the IKEA company\footnotemark[1] uses VR technology to build visual employees to provide services to customers. 
In the energy industry, the E.ON company\footnotemark[2] instructs their substation workers to use VR tools to avoid safety accidents. 
In the education industry, Google Expeditions~\cite{articleGoogle} are used to make immersive education. 
For the tourism industry~\cite{articletourism}, VR technology can facilitate immersive site exploration and interactive tours for remote visitors.

The field of visualization has always played an important role in academic and industrial communities~\cite{unknown,huang2023high,cheng2018colormap}.
As one of the fundamental science and business direction, marine and ocean industry can greatly benefit from the application of VR technology.
Existing visualization applications for marine science are mainly aiming to conduct the scientific research, most of which are focused on certain ocean attributes.
There are many representative visualization applications for ocean related exploration such as i4ocean~\cite{articlei4ocean,lili}, google earth~\cite{yu2012google}, World Wind~\cite{bell2007nasa} and so on.
However, these applications lack comprehensive investigation and development for effective and immersive undersea exploration and user-environment interaction.
Furthermore, an application that empowers people to freely explore the undersea environment with immersive interaction is an important and urgent demand for both ocean education and ocean research.


To bridge this gap, we develop a novel application, Deepsea , for immersive roaming in Meta-ocean with VR devices. In this prototype, we design and model two types of ocean seabed scenes, the Indonesia Ocean and the Pacific Ocean, respectively. Our prototype features our lifelike ocean creature modeling and animation pipeline, realistic ocean scene rendering, and flexible undersea exploration interaction. 
The technical details of our prototype are structured as follows: 
the prototype design and each module details are described in Section 2; 
the visual results, functions in the prototype, and implementation details are presented in Section 3; 
a user study that is designed and conducted for the prototype functions evaluation is reported in Section 4.


\footnotetext[1]{IKEA company: https://virsabi.com/ikea-virtual-reality-for-onboarding-and-training/.} 
\footnotetext[2]{E.ON company: https://www.smart-energy.com/industry-sectors/business-finance-regulation /e-on-trafocusedubstation-workers-with-virtual-reality/.} 

\section{Design and functions}\label{sec1}

This section is organized in the following manner:
We first describe the overview of the design and functions of our prototype. 
Then we focus on two aspects that are critical for undersea roaming, the design of the deep sea environment and the behaviors provided in undersea roaming.

\subsection{Overview}
\begin{figure}[h]
\centering
\includegraphics[width=0.8\textwidth]{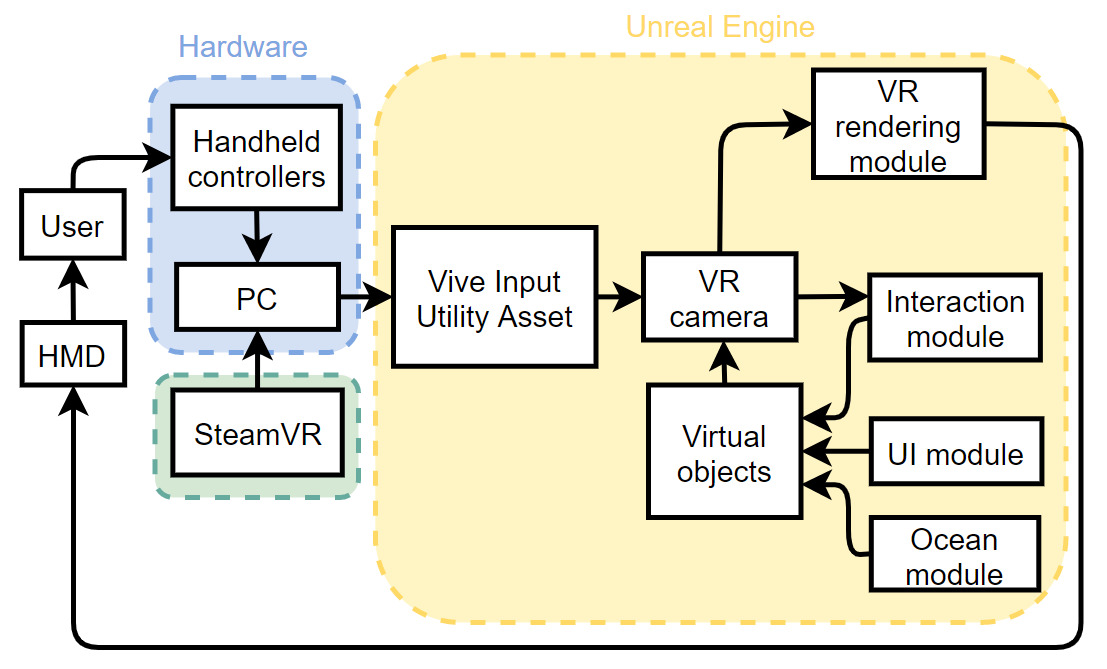}
\centering
\caption{System architecture}
\label{fig:222}
\end{figure}

The architecture of Deepsea is shown in Figure~\ref{fig:222}. 
The hardware includes the personal computer(PC) and the HTC Vive Pro headset.
The HTC Vive Pro headset includes a head-mounted display (HMD), two handheld controllers, and two lighthouses. 
The handheld controllers are used to interact with the graphical user interface (UI) in Deepsea.
The right handheld controller is responsible for emitting light rays to click a virtual button in order to choose the functions.
 The left handheld controller is in charge adjusting the floating and diving motion of the user. 
We use the SteamVR library\cite{murray2020building} to connect the headset and the PC.

In the Unreal Engine, there includes four sub-modules: VR rendering module, interaction module, UI module and ocean module. 
The VR rendering module displays the virtual contents to the users in the VR scene with the help of Vive Input Utility (VIU). 
The UI module provides text and image instructions to the users and allows them to interact with the UI elements in the VR scene with the help of the handheld controllers. 
The interaction module provides interesting knowledge in the VR roaming scene.
The ocean module includes the marine biological models which are modeled according to real marine creature.

\begin{figure}[h]%
\centering
\includegraphics[width=0.7\textwidth]{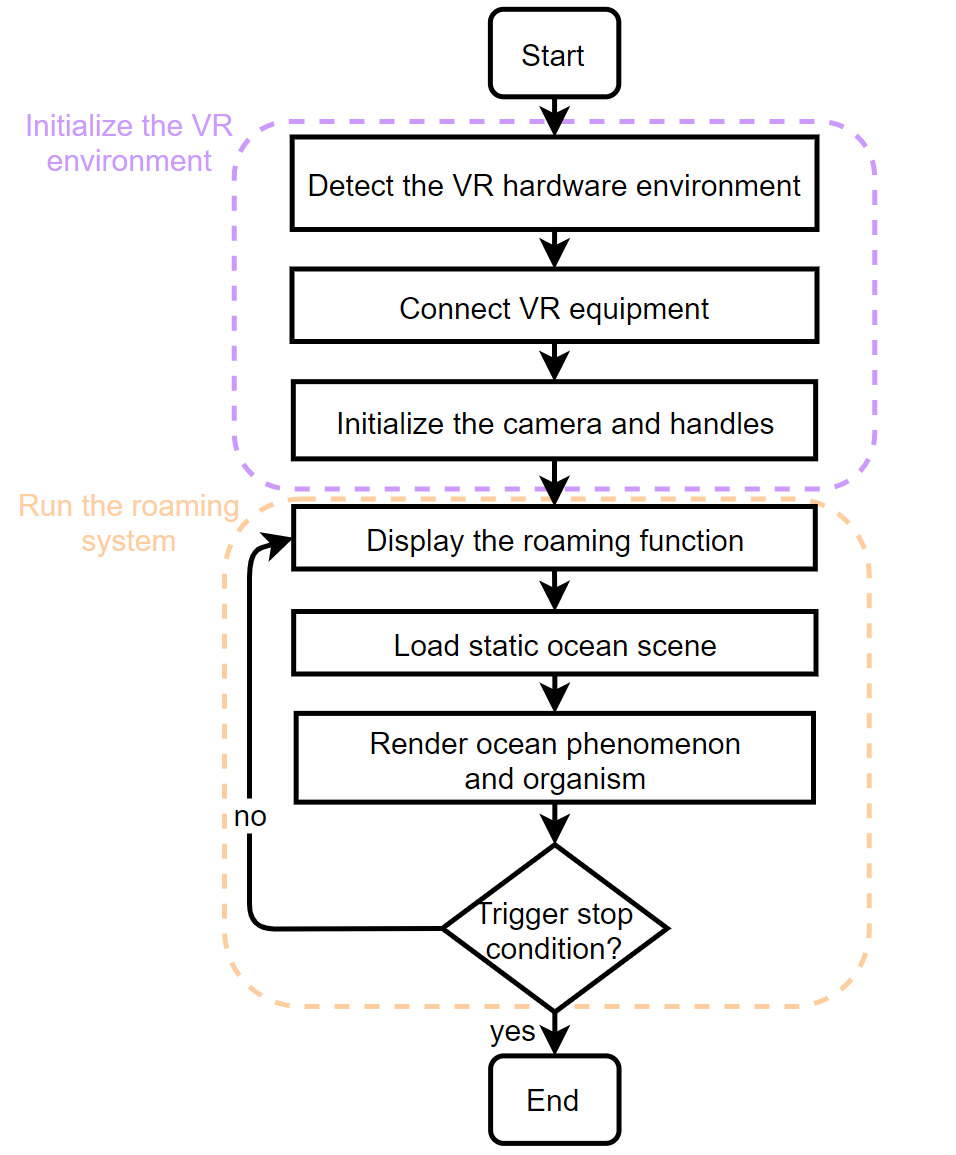}
\caption{Data streaming and control signal flow in Deepsea runtime process}\label{fig:223}
\end{figure}

The flowchart of this prototype for roaming in the undersea is shown in Figure~\ref{fig:223}. 
In the first step, this prototype initializes the VR environment. 
The prototype detects the VR hardware environment and connects the VR equipment including the VR helmet and two handheld controllers. 
After that, the users can enter the VR room where the parameter setting of the camera and controllers can be initialized in a virtual environment. 
In the second step, the prototype for roaming in the undersea begins to run. 
The functions are chosen by users, then the prototype loads the static ocean scene according to the users' choice. 
The ocean phenomena and creatures are rendered to establish the immersive ocean scene.

\subsection{Modeling}

\begin{figure}[h]%
\centering
\includegraphics[width=1\textwidth]{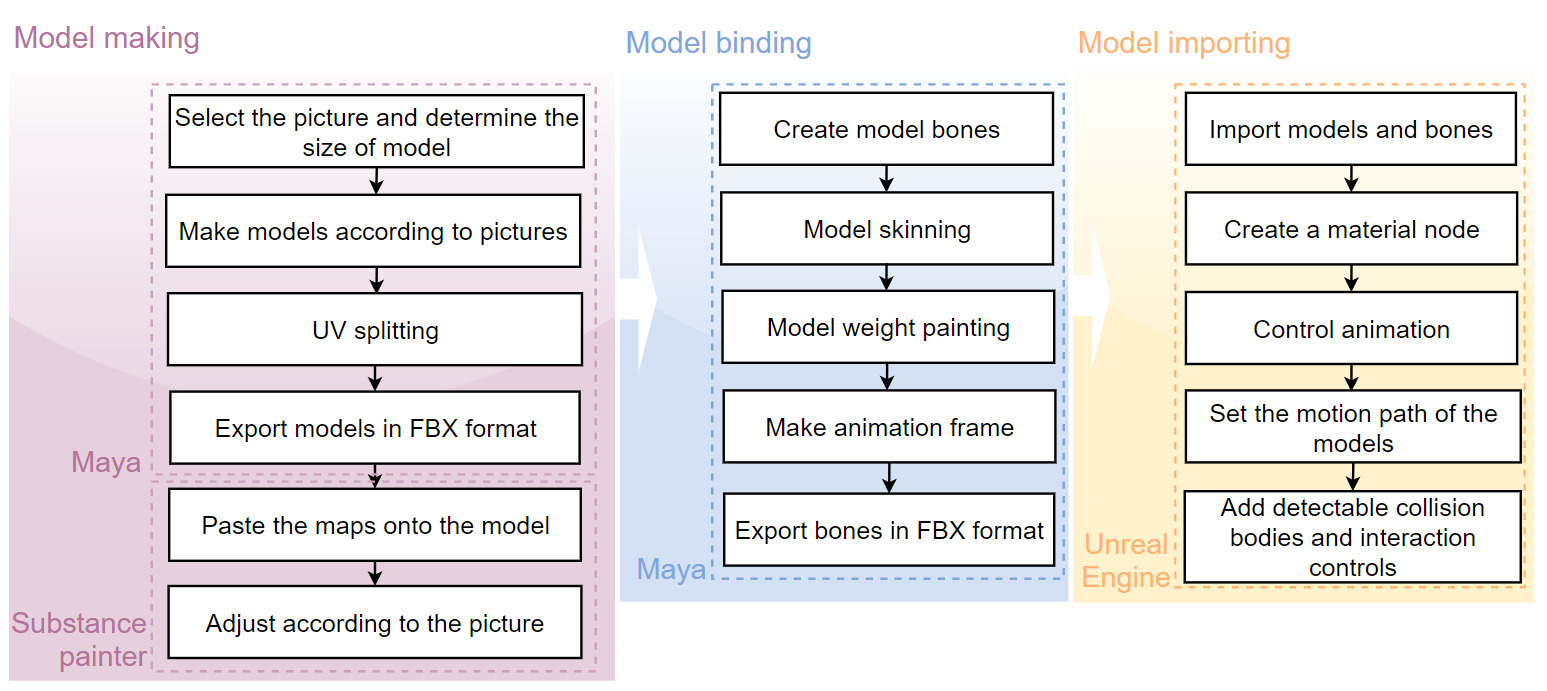}
\caption{Pipeline for ocean creature modeling process.}\label{fig:225}
\end{figure}

For the modeling process, we can divide it into three parts, model making, binding, and importing, as shown in Figure~\ref{fig:225}.
Our model making utilizes two prototype. 
We first select the picture and set the size of the model. Then we implement the UV splitting and export it in FBX format from Maya into Substance painter.
Secondly, the prototype we used is substance painter. We paste the maps which are made in the last part onto the model and adjust the details according to the reference picture.

In model binding, we use Maya to bind models. 
We first create the model bones and implement the skinned Mesh. Then we paint the model weights and make the animation frame. After that, we export the bones in FBX format from Maya.
In model importing, we use the Unreal engine to process the last part. First of all, we import the models and bones built in the above parts into Unreal Engine\cite{karis2013real}. After that, we create the material nodes and set the number and position for the models. Next, we control the animation of models and set the key points to format the motion path of the models. In the last, we add the detectable collision bodies and interaction control for the models.

\begin{figure}[h]%
\centering
\includegraphics[width=1\textwidth]{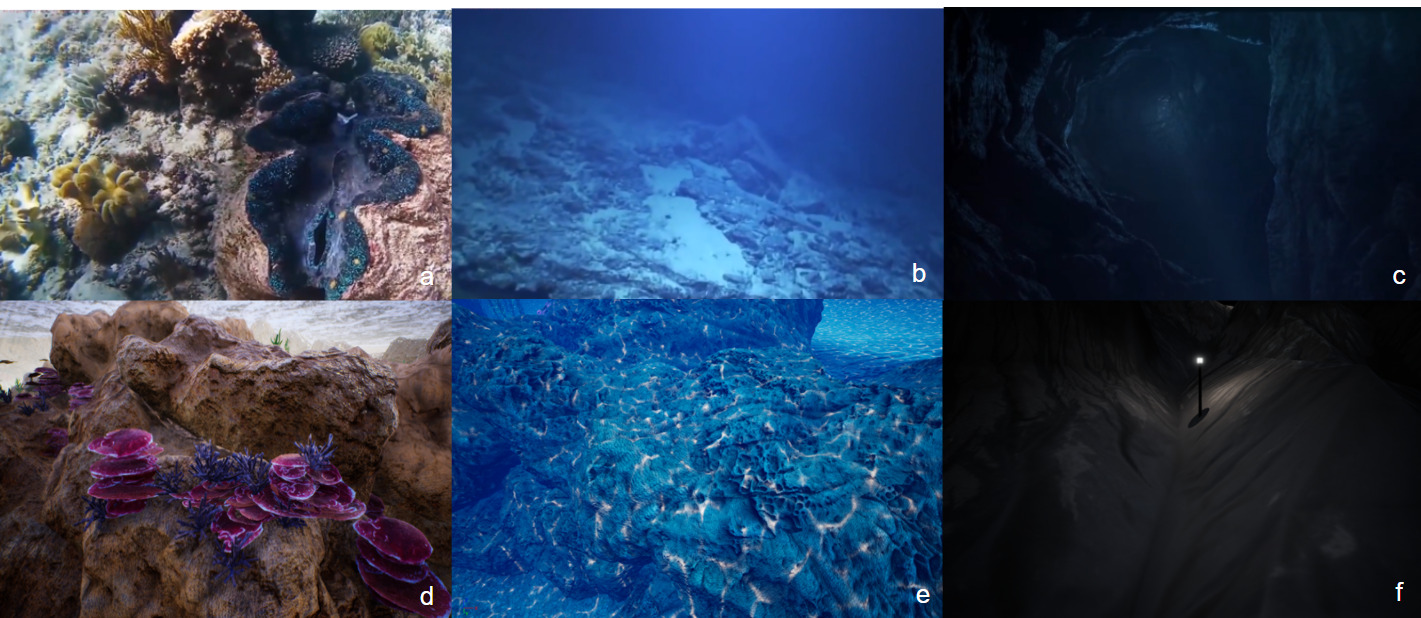}
\caption{ The comparison of the real ocean scene and the models. Figures(a-c) are the real scenes in the seabed while the figures (f-i) are the modeling scenes in the Deepsea. Figures(d-f) are model process with bones.}\label{fig:227}
\end{figure}

\begin{figure}[h]%
\centering
\includegraphics[width=1\textwidth]{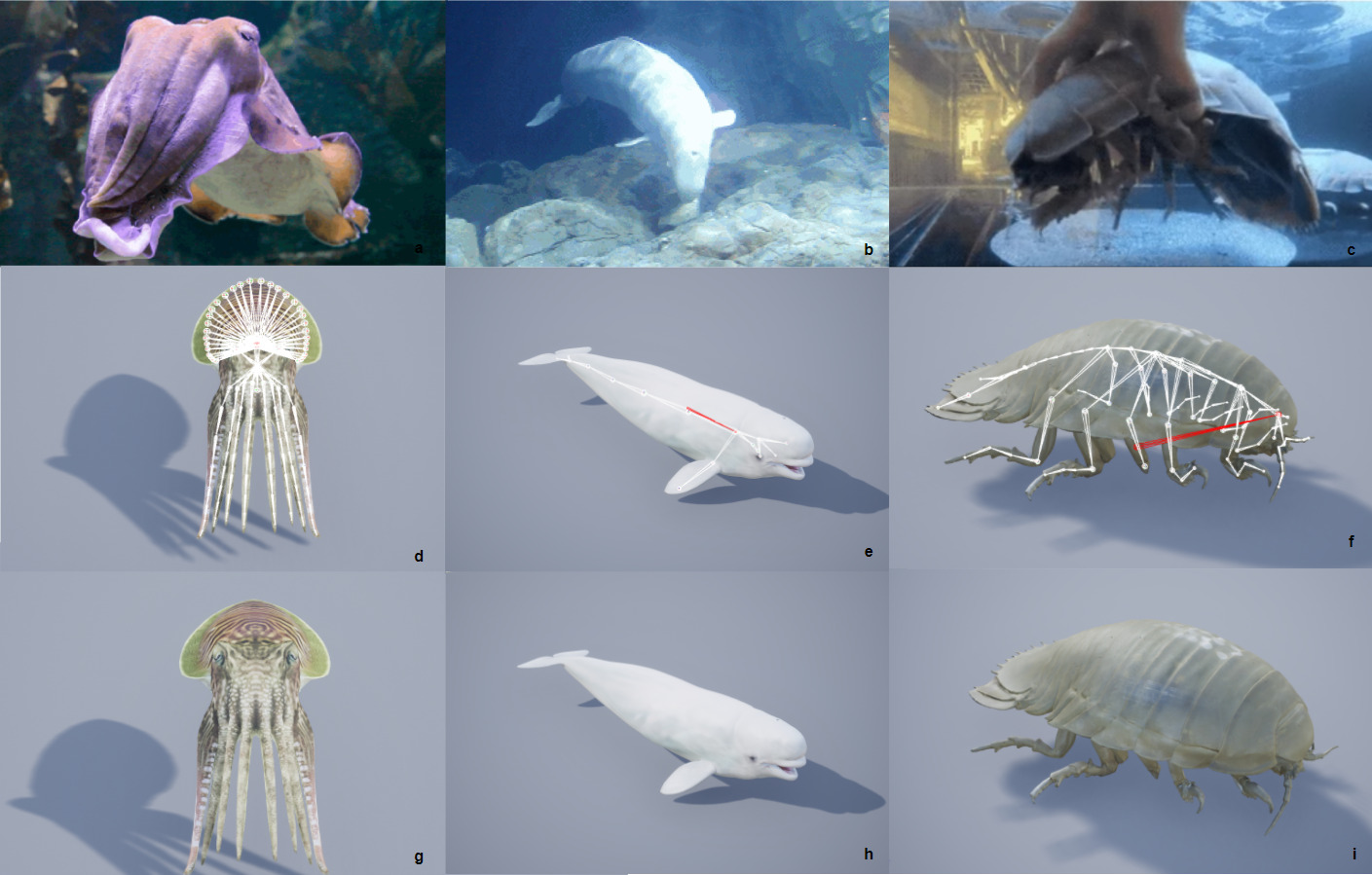}
\caption{The comparison of the real marine creatures and the models}\label{fig:226}
\end{figure}

The models used in the Deepsea can be divided into two types. 
The first is the ocean scene models which can be seen in Figure~\ref{fig:227}. Figures~\ref{fig:227}[a-c] are the real ocean scene while Figures~\ref{fig:227}[d-f] are models used in Deepsea according to the Figures~\ref{fig:227}[a-c]. From them, we can see that the Deepsea has restored the real ocean scene as much as possible.
The second is the marine creature as shown in Figures~\ref{fig:226}.
Figures~\ref{fig:226}[a-c] are the real marine creatures, which are used as objects of reference in the process of modeling. Figures~\ref{fig:226}[d-f] show the bones of marine creature models. Figures~\ref{fig:226}[g-i] are the final models used in the Deepsea.
From Figure~\ref{fig:227} and Figure~\ref{fig:226}, we can find that there also exists a high similarity between the real creature and the visual models, which can reflect the effectiveness of our model processing.

\subsection{Adding behaviors}

\begin{figure}[h]%
\centering
\includegraphics[width=0.8\textwidth]{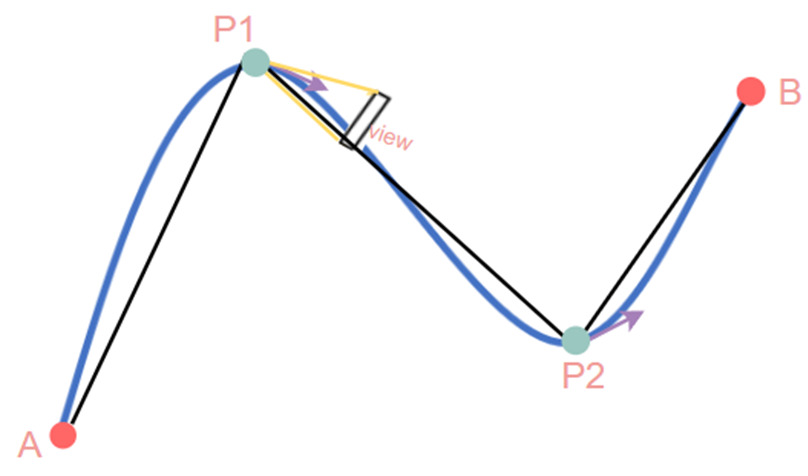}
\caption{Path optimization principle}\label{fig:228}
\end{figure}

For marine creatures, their behaviors are mostly along the key points. 
We first enter a path containing multiple keys and set their longitude, latitude, height, and speed. Then we determine the starting point, the ending point, and the waypoints according to the order of this path.
In the process of path travel, the perspective of the camera is always changed because it will turn to the next node when a node is reached.
As Figure~\ref{fig:228} shows, A is the starting point and B is the ending point. P1 and P2 are the waypoints. 

\begin{figure}[h]%
\centering
\includegraphics[width=0.8\textwidth]{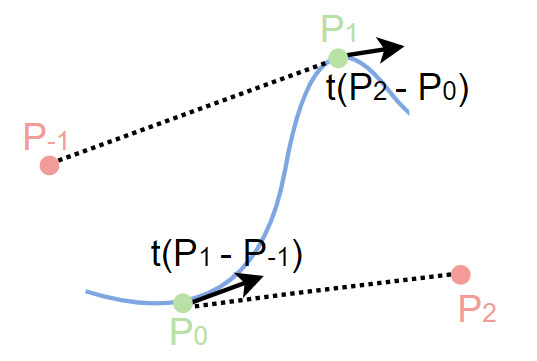}
\caption{Catmull-Rom spline algorithm diagram}\label{cat}
\end{figure}

In this figure, the purple arrow indicates the tangent of the point, and the yellow box marked with view indicates the direction of the camera. The black line represents the traditional polyline path while the blue curve represents the roaming path. From Figure~\ref{fig:228} we can see that the optimization path will bring a better roaming effect for both marine creatures and users.
In this section, we propose a path optimization algorithm based on the Catmull-Rom spline as shown in Figure~\ref{cat}.

As shown in Figure~\ref{cat}, the Catmull-Rom spline needs at least 4 points to control. 
$P_{-1},P_2$ are the starting point and ending point while $P_0, P_1 $ are the control points. The t is a parameter ranging from 0 to 1 to control the parallelism between the curve and the control points, which will determine the fitting shape of this curve.
The Catmull-Rom spline can be defined by Equation~\ref{eq1} to Equation~\ref{eq3}.
\begin{equation}
P(u)=au^3+bu^2+cu+d
\label{eq1}
\end{equation}

\begin{equation}
P(0) = P_{0} , P(1) = P_{1}
\label{eq2}
\end{equation}

\begin{equation}
P^{\prime}(0)=t(P_1 - P_{-1}), 
P^{\prime}(1)=t(P_2-P_0)
\label{eq3}
\end{equation}
%
%
%
%
%
%

\begin{figure}[h]%
\centering
\includegraphics[width=1\textwidth]{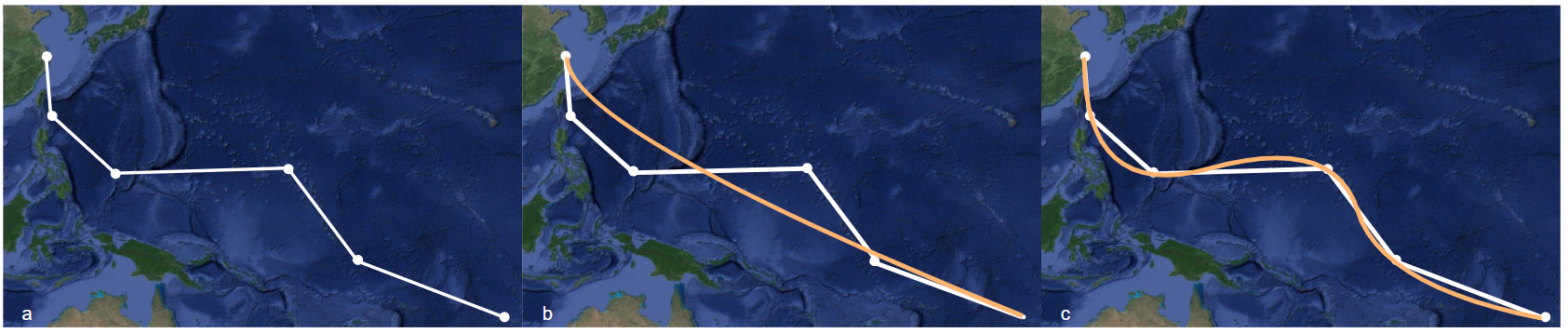}
\caption{Comparison diagram of different paths. a is the original polyline path, b is the Bezier path, and c is the Catmull-Rom path}\label{fig:229}
\end{figure}

To compare the path optimization effect of different curves, we set the six key points to determine different paths according to different curve algorithms. The key points can be seen in Table~\ref{tab1}.
We use the traditional polyline path, the bezier curve path, and the catmull-rom curve path to compare the effect of path optimization. 
As shown in Figure~\ref{fig:229}[a],  we can see that the traditional polyline path has the obvious corner in the connection of key points, which will lead to stiff steering and uneven view change. Compared with that, the paths in Figure~\ref{fig:229}[b] and Figure~\ref{fig:229}[c] both have better smoothness and continuity. It is worth noting that the path generated using Bezier curve did not coincide with key points in Figure~\ref{fig:229}[b]. This is determined by the characteristics of the Bezier curve. The Bezier curve could only be approximated infinitely but could be not pass through key points. Compared with that, the path based on Catmull-Rom could not only be a smooth Path but also pass through all key points as shown in Figure~\ref{fig:229}[c].

 \begin{table}[h]
 \begin{center}
 \begin{minipage}{174pt}
 \caption{3D coordinate values of key points}\label{tab1}%
 \begin{tabular}{@{}llll@{}}
 \toprule
 Key points & Longitude  & Latitude & Height \\
 \midrule
 0    & 121.47   & 31.23  & 10000  \\
 1    & 123.00   & 20.00  & 50000  \\
 2    & 135.00   & 10.00  & 50000  \\
3    & 170.00   & 13.00  & 50000\\
4    & 180.00   & -5.00  & 50000\\
5    & 200.00   & -13.50  & 50000\\
 \botrule
 \end{tabular}
 \end{minipage}
 \end{center}
 \end{table}

\section{Results}\label{sec1}

The Deepsea is developed using the C++ programming language on Unreal Engine\cite{karis2013real} and run on a 3.00 GHz Intel Core i5-12490F processor, 16.0 GB RAM, and 12.0 GB dedicated GPU memory.

%

\begin{figure}[h]%
\centering
\includegraphics[width=0.5\textwidth]{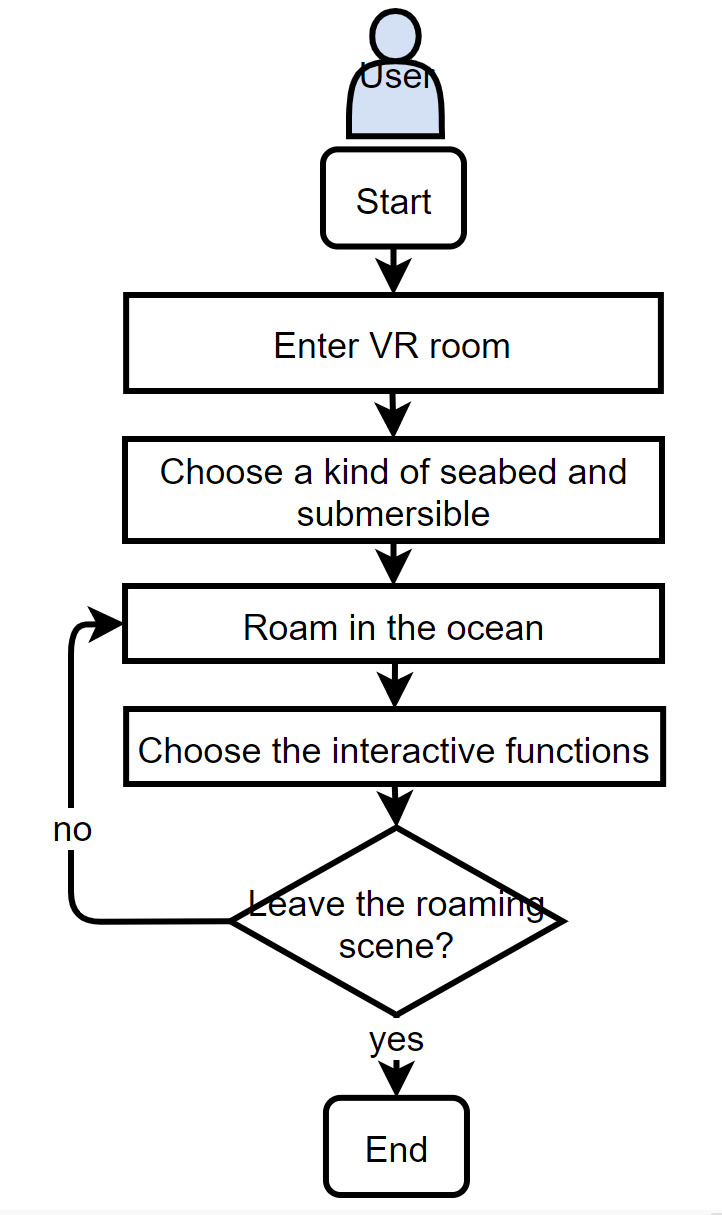}
\caption{Flowchart of roaming in the undersea with Deepsea for users}\label{fig:224}
\end{figure}

\begin{figure}[h]%
\centering
\includegraphics[width=1\textwidth]{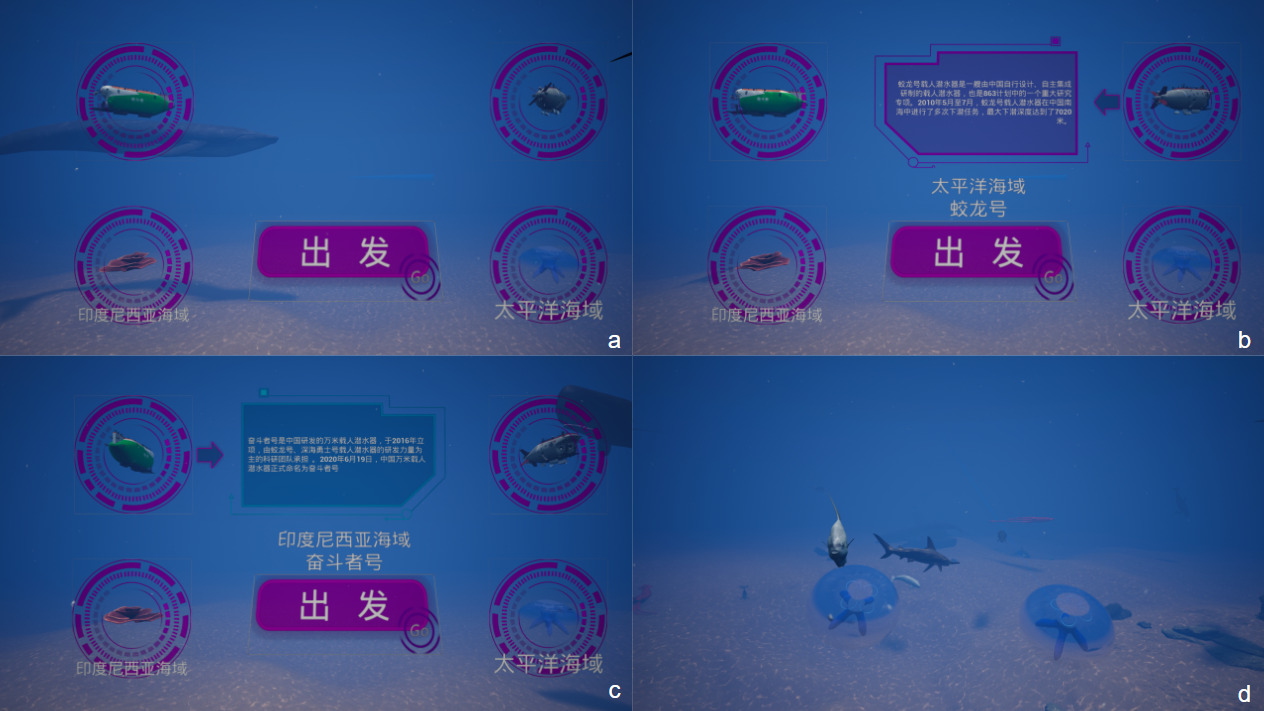}
\caption{Home page of Deepsea}\label{fig321}
\end{figure}

The flowchart of roaming in the Deepsea can be seen in Figure~\ref{fig:224}.
We enter the VR room which is supported by steam and begin to the undersea roaming with the Deepsea.
First of all, we can find the home page as shown in Figure~\ref{fig321}. On this page, we can choose a kind of seabed from the ocean of Indonesia and the Pacific.
The interaction way of this page need to use the control handles. The Deepsea shapes the control handles into jellyfish, and users can emit rays to the menu which can choose the functions.

\begin{figure}[h]%
\centering
\includegraphics[width=1\textwidth]{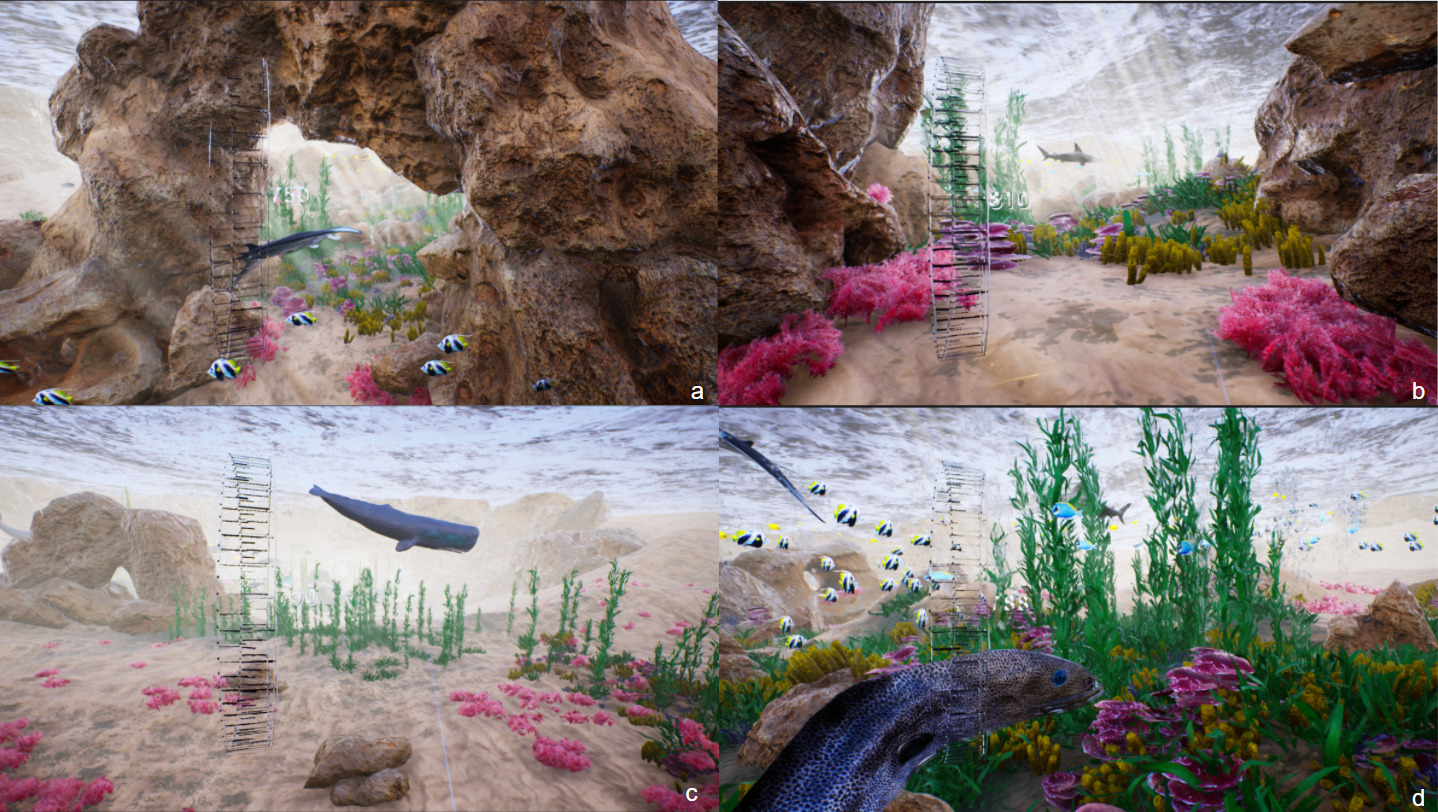}
\caption{The Indonesia ocean in Deepsea}\label{fig:322}
\end{figure}

\begin{figure}[h]%
\centering
\includegraphics[width=1\textwidth]{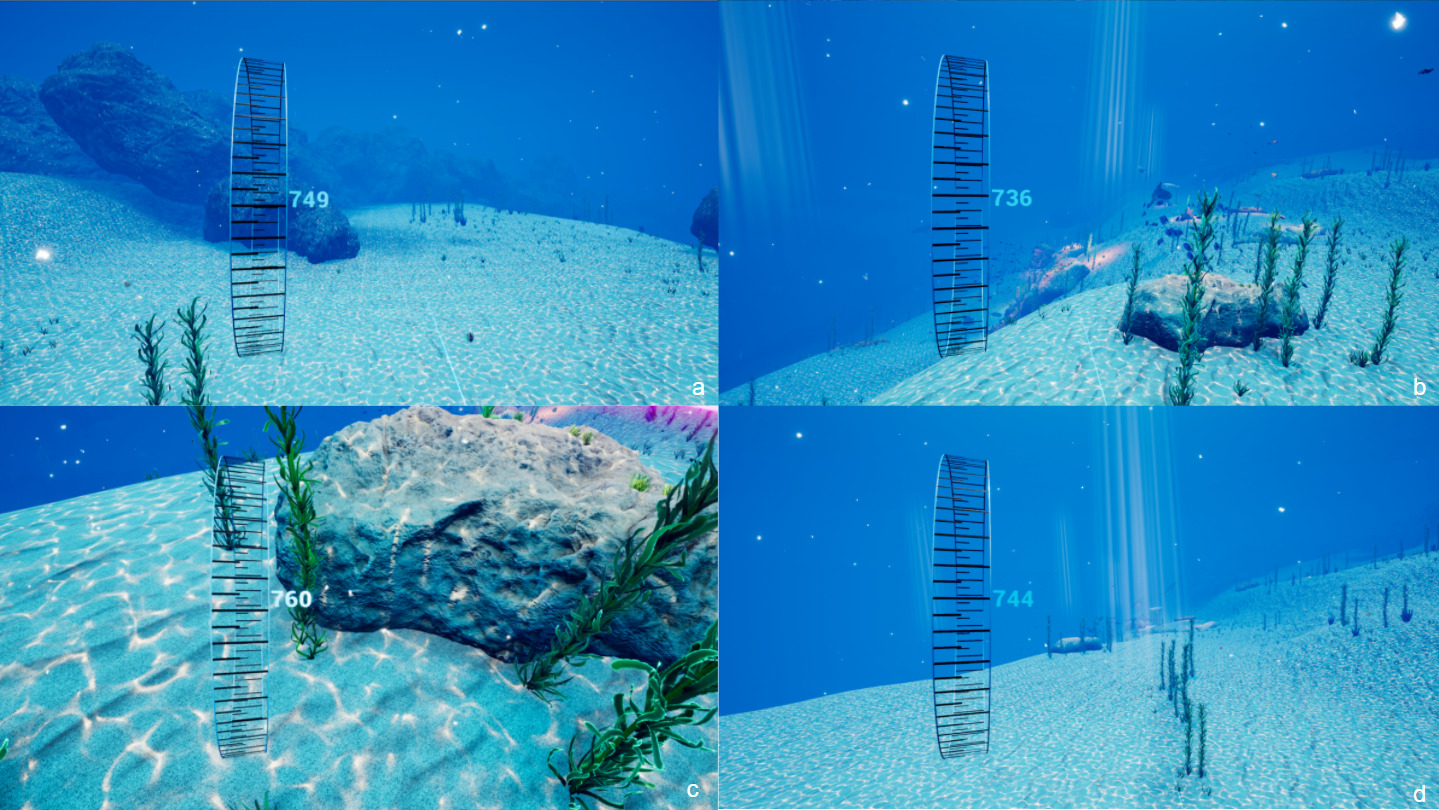}
\caption{The Pacific ocean in Deepsea}\label{fig:326}
\end{figure}

The second step is roaming the undersea. The Deepsea provides two sea scenes which are the Indonesia ocean and the Pacific ocean as shown in Figure~\ref{fig:322} and Figure~\ref{fig:326}.
Users can use the left handle to control the direction of roaming and use the right handle to emit ray determining functions. 
We set different models according to the real sea area and strive to provide a kind of immersive and real roaming experience to users.

\begin{figure}[h]%
\centering
\includegraphics[width=1\textwidth]{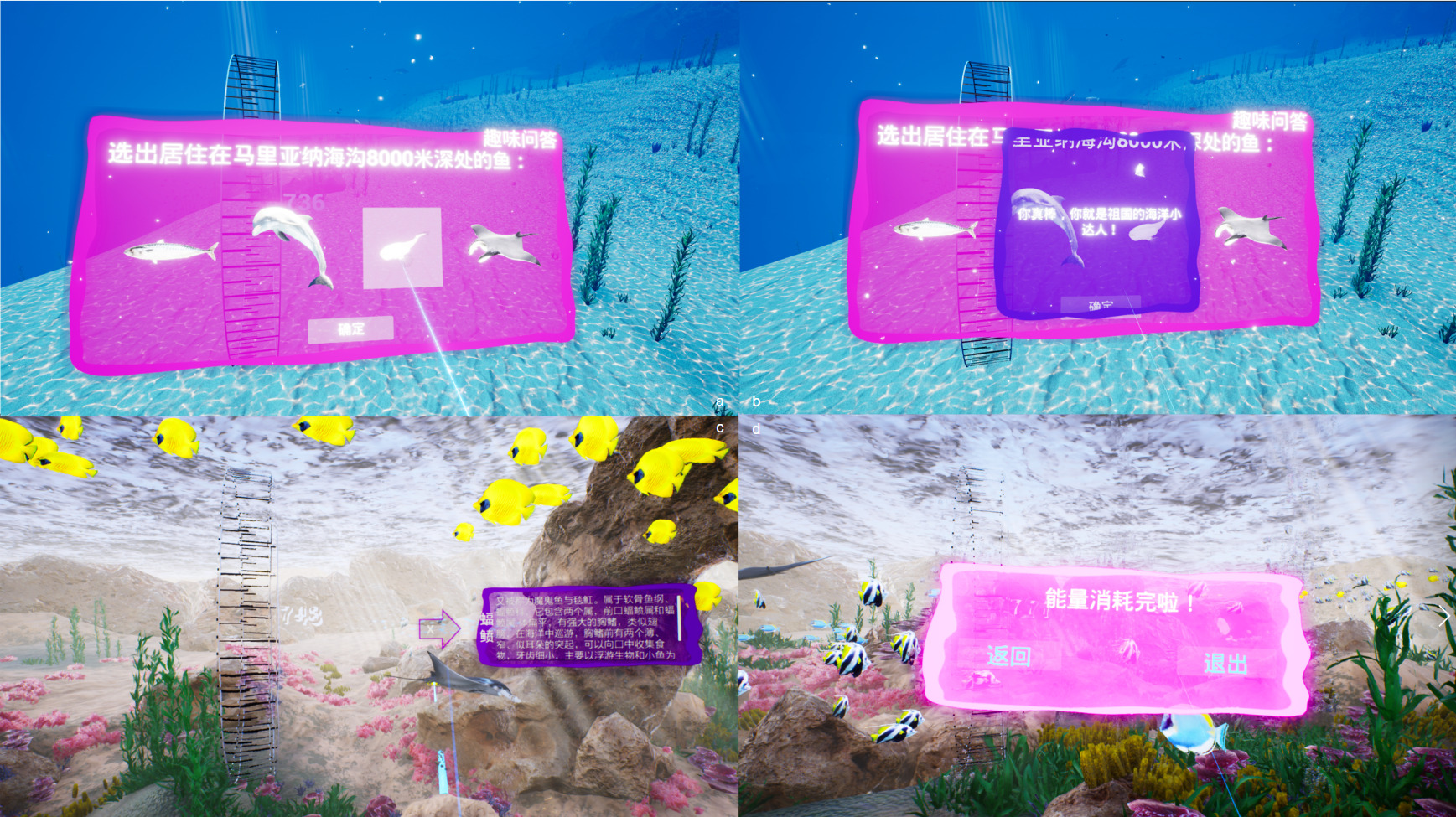}
\caption{The interaction in Deepsea}\label{fig:325}
\end{figure}

In the process of roaming, we provide some interactions to users as shown in Figure~\ref{fig:325}. 
There are some interesting questions about the current sea area. Users can choose answers by emitting rays using the right control handle as shown in Figure~\ref{fig:325}[a-b]. 
Besides, users can select marine creatures by emitting rays while they are roaming in undersea. If the ray exactly selects the target creature, there are text boxes occuring and showing the detailed knowledge about that creature for users as shown in Figure~\ref{fig:325}[c]. In order to add the immersion, we also add the auto-voice system for this function.
Finally, there also have exit options. For example, if the roaming time of users is more than five minutes, a prompt box will occur which will remind users that the energy is exhausted and they need to exit as shown in Figure~\ref{fig:325}[d].

\clearpage
\section{User Study}\label{sec1}

This section is organized as follows:
First, we design a few roaming tasks and two questionnaires to examine the users’ enjoyment and engagement. 
Second, we report the time of completing roaming tasks and the total scores of each questionnaire. 
Finally, we examine the relationships between the time to complete the roaming task and the performance of users based on the above-mentioned evaluation.

\subsection{Participants}

We invite 50 healthy participants to take part in our user study. The participants involve 22 males and 28 females, whose mean age is 25.2 years old. 
Besides, we also set strict healthy exclusion criteria in order to avoid any discomfort in using the VR device, i.e., we exclude people with significant cognitive dysfunction, previous neurological illness psychiatric, heart condition, other serious medical conditions, significant mobility restrictions, seizures, or with interfering medical devices. We also exclude pregnant woman and elderly individuals. 

\subsection{Tasks and questionnaires}

We set roaming tasks in the seabed scene for participants in order to test the simplicity and effectiveness of Deepsea during the roaming. We set a few key points in the scene as a roaming path and record the time used in roaming, the total number of collisions, and the accuracy with the UI rays. For accuracy, we record the total number and the effective number of emitting rays to UI, then we can calculate the accuracy using the above numbers. 
The results can be seen in Table~\ref{tab1a}. We make statistics of results and calculate the mean value and the standard deviation(SD). 
From Table~\ref{tab1a} we can find that the time used in roaming and the number of collisions are lower than the numerical value that we expected. Besides, the lower SD also reflects that there exists strong stability in the experience of roaming with Deepsea.
Besides, the third variable, accuracy with the UI rays, also shows the effectiveness of UI module.

\begin{table}[h]
\begin{center}
\begin{minipage}{170pt}
\caption{Results of roaming tasks}\label{tab1a}%
\begin{tabular}{@{}llll@{}}
\toprule
Variable &  Mean   \\
\midrule
Time used in roaming (s)        &  213.4    \\
Number of collisions         &  2.3        \\
Accuracy with the UI rays    &  0.75    \\

\botrule
\end{tabular}

\end{minipage}
\end{center}
\end{table}

\begin{table}[h]
\begin{center}
\begin{minipage}{230pt}
\caption{Statements used to quantify users’ enjoyment}\label{tab1}%
\begin{tabular}{@{}llll@{}}
\toprule
Variable & Statement to be graded  & Mark  & Mean   \\
\midrule
             & I was interested in tasks    & 1-5  &4.6   \\
             & I found it impressive    & 1-5 & 4.3  \\
 Enjoyment   & I felt skillful      & 1-5& 3.2\\
             & I felt completely absorbed        & 1-5& 4.5\\
             & I found it easy to use      &1-5& 4.1\\
            
             \botrule
 \pmb{Sum   }         &                         &  \pmb{5-25 }&  \pmb{ 20.7 }   \\

\botrule

\end{tabular}

\end{minipage}
\end{center}
\end{table}

There are two questionnaires designed to evaluate the enjoyment and engagement of participants.
The first questionnaire consists of nine statements, which mainly focus on the enjoyment of participants.
The participant could choose a level of their agreement with using a five-point scale (not at all, slightly, moderately, fairly, and extremely). Responses of participants were associated with a score ranging from 1 to 5.
The second questionnaire is made up of six questions that mainly focus on the engagement of participants.  
The participants were asked to use an "X" to fill in the appropriate box of a five-point scale according to their experience. 
For all questionnaires, the participants were also prompted to consider the entire scale to answer the questions independently, and not skip questions or return to a previous question to change their answers.
The results of the questionnaires can be seen in Table \ref{tab1} and Table \ref{tab2}.

According to the feedback from the participants, most people positively think they experience a strong sense of immersion. Besides, participants agree that they can focus on the tasks given the interaction functions. On the other hand, participants hold various opinions regarding how natural the interaction in this prototype is. The results of the questionnaires can be seen in
Table \ref{tab1} and Table \ref{tab2}. Overall, we can find that the Deepsea can bring enjoyment and engagement for participants


\begin{table}[h]
\begin{minipage}{274pt}
\caption{Questions used to quantify users’ engagement}\label{tab2}%
\begin{tabular}{@{}llll@{}}
\toprule
Variable & Question to be graded   & Mark & Mean\\
\midrule
             & How involved were you in the virtual environment experience?  &  1-5  & 3.8 \\
             
             & Please rate your sense of being in the Deepsea.&  1-5  & 4.1\\
             
Engagement              &How natural did your interactions with the Deepsea seem? &  1-5 & 3.6 \\

             
             & \begin{tabular}[c]{@{}l@{}}How much did you feel the handheld controllers were your own\\\quad hands? \end{tabular} & 1-5  & 3.7\\
             
             & \begin{tabular}[c]{@{}l@{}}How well could you concentrate on the assigned tasks rather \\\quad than on the mechanisms used to perform those tasks? \end{tabular}  &  1-5 & 3.7\\

\botrule
 \pmb{Sum   } &  & \pmb{5-25}  & \pmb{18.9} \\
\botrule
\end{tabular}

\end{minipage}
\end{table}

\subsection{Hypotheses and statistical analysis}
In this section, we have conducted an analysis to explore the connection between the enjoyment and engagement scores of users with using a questionnaire-based evaluation. 
Subsequently, we investigated whether the engagement score exhibited any correlation with two specific task-related variables: the time taken to complete the task, the total number of collisions and accuracy with the UI rays.

\subsubsection{Hypotheses}
We set the following hypotheses to analysis the correlation of the Deepsea and the tasks.

H1: Participants’ enjoyment and engagement scores are positively related.

H2: Participants’ performance and their engagement score are related.

Hypotheses H1 and H2 are investigated through the completion of tasks.
Based on previous studies on serious games, we expect to find a correlation between enjoyment and engagement. 
Thus, we set a positive hypothesis like H1.
If such a correlation exists, it will provide valuable insights for future design interventions in the VR platform, aiming to enhance enjoyment and subsequently improve engagement.
In H2, our hypothesis posit that there will be an association between participants' performance measures, such as the total time taken to complete the experiment, the total number of collisions, and accuracy with the UI rays, and their engagement score.
We anticipat that if such an association is observed, it will demonstrate the potential of using VR for operations management training.

\subsubsection{Analysis of engagement, enjoyment and performance} 

We first employ the Kolmogorov–Smirnov (K-S) test to confirm whether the variable data is a normal distribution. The results are as follows:
The engagement scores (D=0.107, p=0.2), enjoyment scores (D=0.124, p=0.054), the time taken to complete the task(D=.056 ,p=0.2 ) and the accuracy with the UI rays (D=0.106, p=0.2) are all normally distributed. 
And the total number of collisions (D=0.287, p=0) is not following a normal distribution.

\begin{figure}[h]%
\centering
\includegraphics[width=0.8\textwidth]{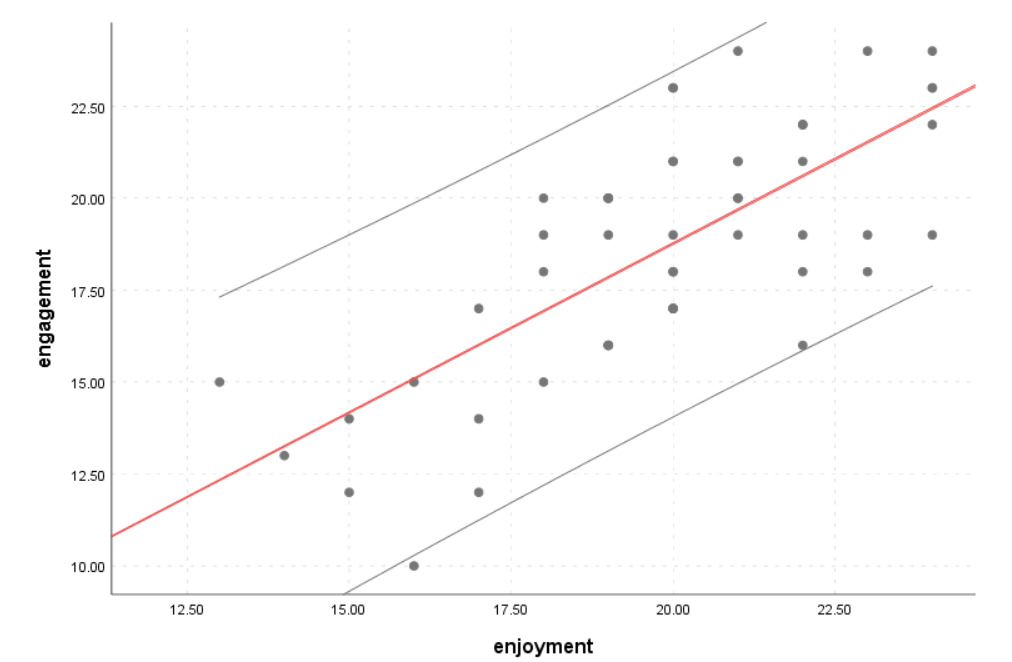}
\caption{Scatter plot of enjoyment and engagement scores. 
Points correspond to individual realizations and area between two gray lines represents the
95 confidence interval of the linear regression.}\label{figenjoy}
\end{figure}

Then, for each Hypotheses we will choose the appropriate algorithms to explore the correlation between variables.
To test H1, we employ the Pearson correlation coefficient to determine the correlation between enjoyment scores and engagement scores (r = 0.732, p = 0.0). The scatter plot in Figure~\ref{figenjoy} provides visual evidence of this correlation. 
The statistically significant correlation coefficient confirms the existence of a strong association between enjoyment scores and engagement scores, thereby validating hypothesis H1.

\begin{figure}[h]%
\centering
\includegraphics[width=0.8\textwidth]{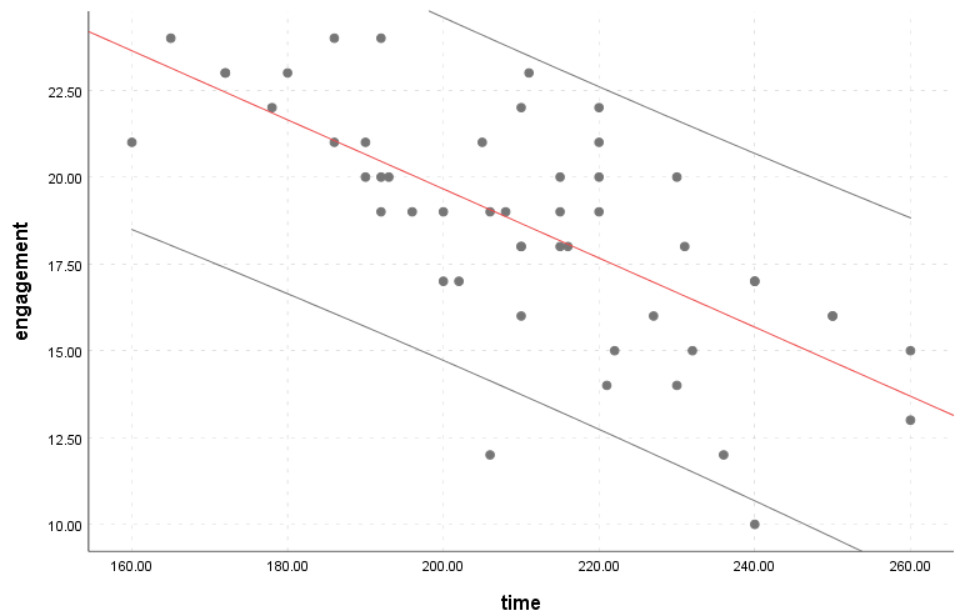}
\caption{Scatter plot of time and engagement scores. 
Points correspond to individual realizations and area between two gray lines represents the
95 confidence interval of the linear regression. }\label{fig_time}
\end{figure}

\begin{figure}[h]%
\centering
\includegraphics[width=0.8\textwidth]{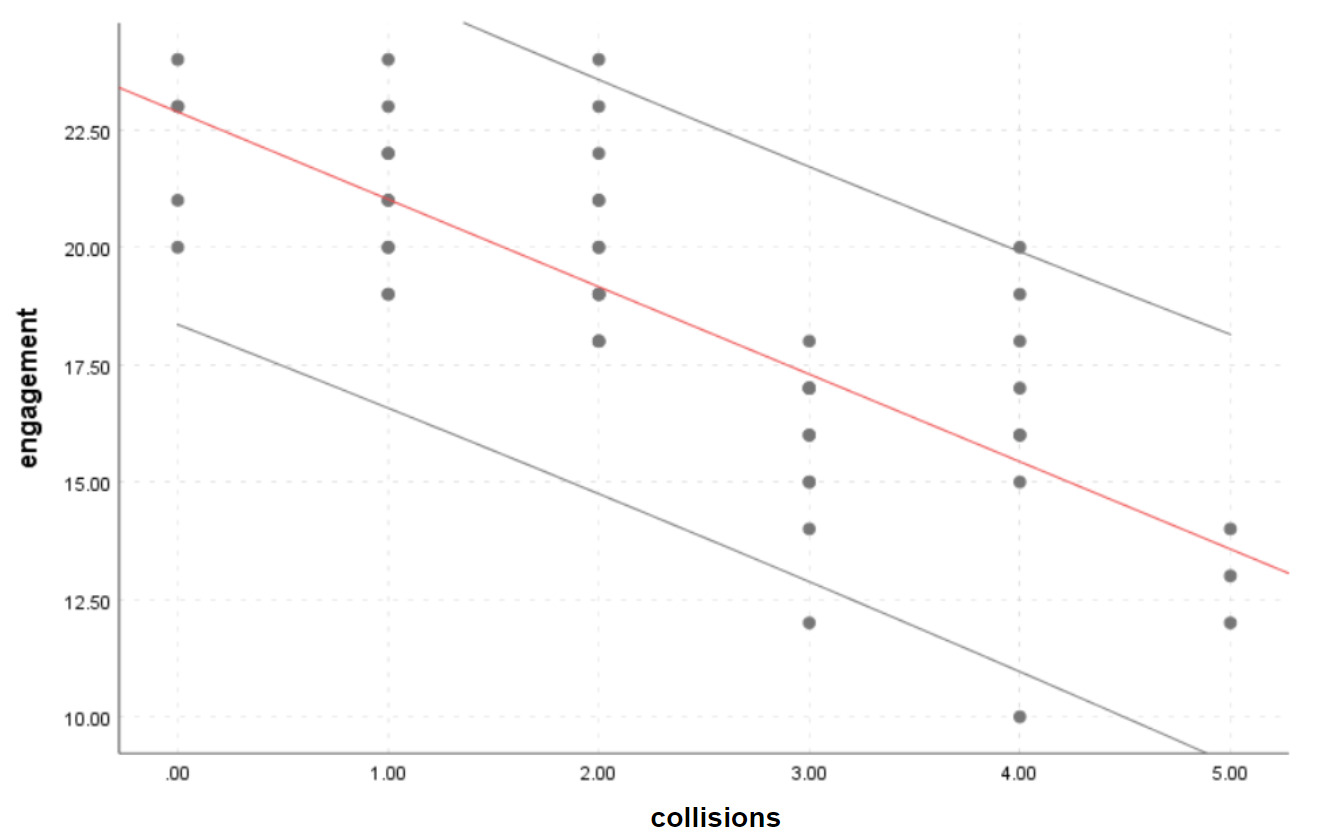}
\caption{Scatter plot of collision and engagement scores. 
Points correspond to individual realizations and area between two gray lines represents the
95 confidence interval of the linear regression. }\label{fig_collsion}
\end{figure}

\begin{figure}[h]%
\centering
\includegraphics[width=0.8\textwidth]{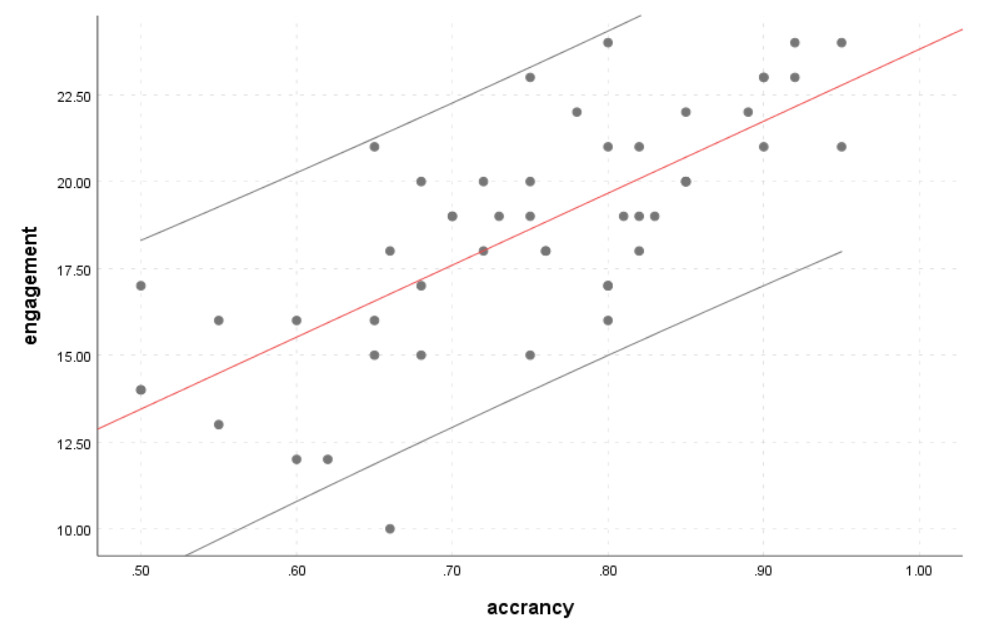}
\caption{Scatter plot of accrancy and engagement scores. 
Points correspond to individual realizations and area between two gray lines represents the
95 confidence interval of the linear regression. }\label{fig_accrancy}
\end{figure}

To test H2, we employ the Pearson correlation coefficient to determine the correlation between engagement scores and the total number of collisions (r = -0.7, p = 0), as well as the correlation between engagement scores and accuracy with the UI rays (r = 0.737, p = 0). 
These correlations are based on the assumption that the data followed a normal distribution. 
The statistically significant correlation coefficients confirm a relatively strong association between engagement scores and both the total number of collisions and accuracy with the UI rays. The scatter plots in Figure~\ref{fig_time} and Figure~\ref{fig_accrancy} provide visual representations of these correlations.
Since the total number of collisions did not follow a normal distribution, we use the Spearman correlation coefficient to determine the correlation between collisions and engagement scores (r = -0.795, p = 0, Figure~\ref{fig_collsion}). This analysis has confirmed the existence of a correlation between collisions and engagement scores.
Overall, our findings align with our expectations and support hypothesis H2.





~\\
~\\
In this section, we have conducted a user study to gather feedback and insights from users regarding their experience with Deepsea. 
Our primary focus is to examine the relationship between enjoyment and engagement scores, which serve as indicators of user evaluation and interaction within the Deepsea platform. Figure~\ref{figenjoy} visually demonstrates that higher levels of proficiency in Deepsea operations are associated with increased enjoyment. This finding underscores the importance of improving user proficiency in operations to enhance their overall perception of Deepsea.

We first record various aspects of users' operations in Deepsea, including the total time taken to complete the experiment, the total number of collisions, and accuracy with the UI rays. 
Then we proceed to analyze the correlation between these variables and the users' engagement scores using either Pearson or Spearman correlation analyses. 
Notably, Figure~\ref{fig_time} to Figure~\ref{fig_accrancy} visually illustrate the significant relationships observed between engagement scores and these operational variables.

Based on these findings, future efforts will focus on providing prototype pre-training for users and improving the overall user experience. 
This will include the creation of pre-training tutorials, user-friendly prototype enhancements, optimization of collision detection using artificial intelligence algorithms, and offering hardware improvement suggestions to hardware providers. 
These initiatives aim to further enhance user proficiency and interaction smoothness within Deepsea.




\section{Summary and conclusion}\label{sec4}

In this paper, we review the relevant technological development of the Meta-ocean and realize the importance of building the visual environment using VR technology.
Then we design and build our own visual environment, Deepsea, which can provide an immersive experience for users when they explore the Meta-ocean.
We model the ocean scenes and use the Catmull-Rom spine to optimize the roaming path in order to increase immersion.
Finally, we conducted a user study to investigate users' experiences with Deepsea and analyze potential future directions.

Metaverse has already become a hot trend in various fields while ocean exploration has also attracted attention from academics and governments. The Meta-ocean, as an extension of the above topics, will certainly
evolve into a key issue and change the life of humans.
Using VR technology, we propose the Deepsea to make a tentative exploration of the Meta-ocean. This prototype has given an immersive and delightful experience for users.
In future work, we will continue to optimize the interaction module and model more ocean scenes to build more immersive Meta-ocean applications. The Deepsea will be a useful tool for exploring the Meta-ocean.

%
%
%

\bibliography{sn-bibliography}



\end{document}